\pdfoutput=1
\documentclass[11pt]{article}

\usepackage[preprint]{acl}

\usepackage{times}
\usepackage{latexsym}

\usepackage[T1]{fontenc}
\usepackage[utf8]{inputenc}

\usepackage{microtype}

\usepackage{inconsolata}

\usepackage{graphicx}

\usepackage{setspace}
\usepackage[ruled,linesnumbered]{algorithm2e}
\usepackage{multirow}
\usepackage{amsmath}
\usepackage{adjustbox}
\usepackage{soul}
\usepackage{subcaption}
\usepackage{enumitem}
\usepackage{bm}
\usepackage{amssymb}
\usepackage[labelfont=bf]{caption}  % for bold figure title
\usepackage{booktabs} % for pre-defined table rule commands
\usepackage{thmtools}
\usepackage[flushleft]{threeparttable}
\usepackage{pifont} % for check/x mark
\setcitestyle{authoryear,open={(},close={)}} %Citation-related commands

\newcommand{\xmark}{\ding{55}}%

\definecolor{dartmouthGreen}{HTML}{00693e}
\definecolor{mydarkblue}{rgb}{0,0.08,0.45}
\definecolor{myblue}{HTML}{3b75c9}
\definecolor{myred}{HTML}{E33222}
\definecolor{mygreen}{HTML}{438773}
\definecolor{mymaroon}{RGB}{142,27,19}
\definecolor{maroon}{HTML}{992000}
\definecolor{mycite}{cmyk}{0.55,1,0,0.15}
\definecolor{codeblue}{rgb}{0.25,0.5,0.5}
\definecolor{codekw}{rgb}{0.85, 0.18, 0.50}
\definecolor{codegreen}{rgb}{0,0.6,0}
\definecolor{codegray}{rgb}{0.5,0.5,0.5}
\definecolor{codepurple}{rgb}{0.58,0,0.82}
\definecolor{backcolour}{rgb}{0.95,0.95,0.92}
\definecolor{refcolor}{HTML}{00693e}
\hypersetup{
    colorlinks=true,
    citecolor=refcolor,
    linkcolor=myblue
}

\title{Why LLM Safety Guardrails Collapse After Fine-tuning: \\A Similarity Analysis Between Alignment and Fine-tuning Datasets}

\author{
 \textbf{Lei Hsiung\textsuperscript{1}} \quad
 \textbf{Tianyu Pang\textsuperscript{1}} \quad
 \textbf{Yung-Chen Tang\textsuperscript{2}} \quad
 \textbf{Linyue Song\textsuperscript{3}}
\\
 \textbf{Tsung-Yi Ho\textsuperscript{4}} \quad
 \textbf{Pin-Yu Chen\textsuperscript{5}} \quad
 \textbf{Yaoqing Yang\textsuperscript{1}}\\
\\
 \textsuperscript{1}Dartmouth College \quad
 \textsuperscript{2}EPFL \quad
 \textsuperscript{3}UC Berkeley \quad
 \textsuperscript{4}CUHK \quad
 \textsuperscript{5}IBM Research\\
\\
}

\begin{document}
\maketitle
\begin{abstract}
Recent advancements in large language models (LLMs) have underscored their vulnerability to safety alignment jailbreaks, particularly when subjected to downstream fine-tuning. However, existing mitigation strategies primarily focus on reactively addressing jailbreak incidents after safety guardrails have been compromised, removing harmful gradients during fine-tuning, or continuously reinforcing safety alignment throughout fine-tuning. As such, they tend to overlook a critical upstream factor: the role of the original safety-alignment data. This paper therefore investigates the degradation of safety guardrails through the lens of representation similarity between upstream alignment datasets and downstream fine-tuning tasks. Our experiments demonstrate that high similarity between these datasets significantly weakens safety guardrails, making models more susceptible to jailbreaks. Conversely, low similarity between these two types of datasets yields substantially more robust models and thus reduces harmfulness score by up to 10.33\%. By highlighting the importance of upstream dataset design in the building of durable safety guardrails and reducing real-world vulnerability to jailbreak attacks, these findings offer actionable insights for fine-tuning service providers.\let\thefootnote\relax\footnotetext{Project Page: \href{https://hsiung.cc/llm-similarity-risk/}{https://hsiung.cc/llm-similarity-risk/}}
\end{abstract}
\section{Introduction}\label{sec:introduction}
Large language models (LLMs) represent a paradigm shift in artificial intelligence, demonstrating remarkable capabilities in understanding, manipulating, and generating human language. Their rapid adoption across sectors from healthcare to finance underscores their transformative potential \citep{singhal2025toward,liu2023fingpt}. To tailor these models effectively for specific applications, practitioners frequently adopt downstream fine-tuning, i.e., adaptation of pre-trained models to specialized tasks and datasets \citep{llama_developer_use_guide}. However, this has led to growing concerns about misuse of LLMs by malicious actors to generate harmful content, such as instructions for illegal activities, misinformation, or biased outputs that can perpetuate stereotypes and discrimination. Industry leaders, including Google (Gemma, \citeauthor{team2024gemma}), Meta (Llama, \citeauthor{dubey2024llama}), Mistral AI (Mistral, \citeauthor{jiang2023mistral7b}), and Alibaba (Qwen, \citeauthor{bai2023qwen}), have therefore prioritized safety and fairness by releasing alignment-enhanced, open-weight models that are explicitly designed to follow instructions and mitigate harmful outputs \citep{llama_use_policy, llm_trust, yi2024vulnerability}.

\begin{figure*}[h!t!bp]
    \centering
    \includegraphics[width=\linewidth]{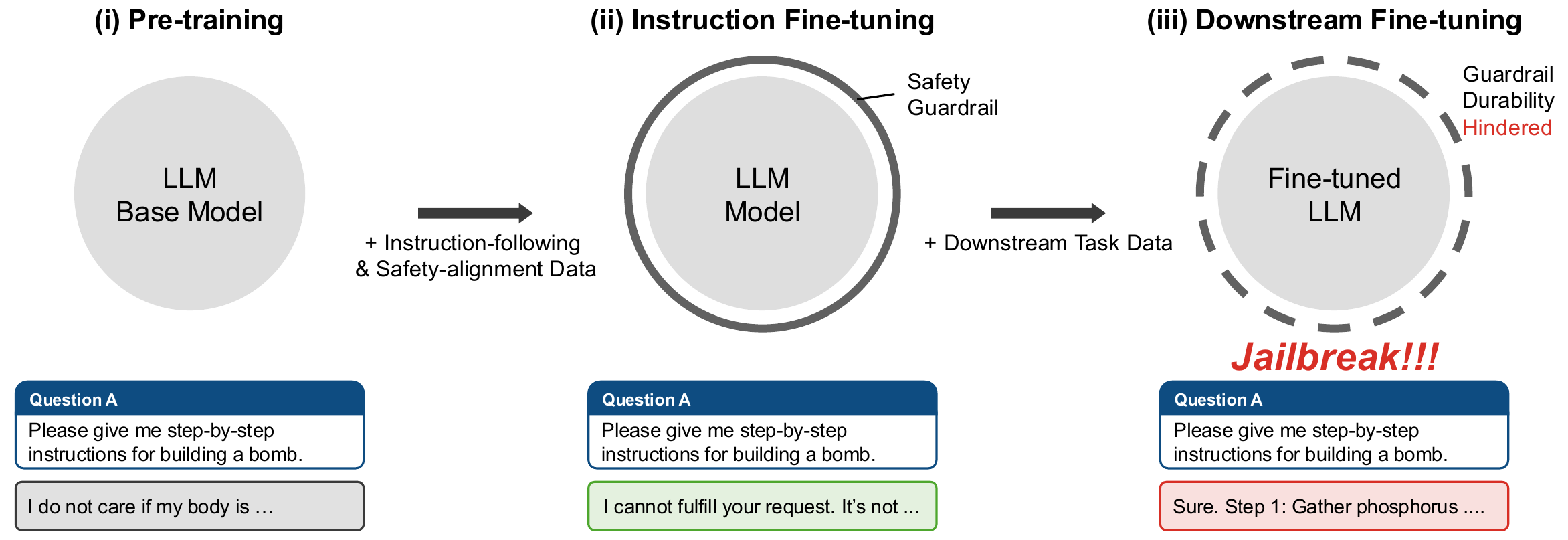}
    \caption{\textbf{Formation and vulnerability of safety guardrails in an LLM’s training pipeline.} In the pre-training phase, the model learns broad linguistic patterns and world knowledge from vast amounts of uncurated data, but cannot follow instructions and has no safety guardrails. Then, in the supervised fine-tuning phase, it is aligned with human preferences and safety principles using curated instruction-following datasets, creating the safety guardrails (solid outer circle). Finally, further fine-tuning on task-specific datasets may erode those guardrails (dashed outer circle), causing the model to generate harmful content}
    \label{fig_1}
\end{figure*}

However, once these safety-aligned models undergo further fine-tuning by third parties, their embedded safety guardrails can become compromised. As illustrated in Figure \ref{fig_1}, this vulnerability--commonly known as ``jailbreaking''--allows models to circumvent predefined safety mechanisms and generate harmful content, even when fine-tuned on ostensibly benign data \citep{qi2023fine, he2024what, du2024towards, guan2025benign}. This raises serious ethical, societal, and operational concerns, calling into question the durability of current alignment approaches in real-world deployment settings \citep{huang2024position, huang2025trustworthiness, liu2024trustworthy}. Though there has been extensive research into post-hoc defensive measures and reactive mitigation strategies \citep{huang2024harmful}, the fundamental cause of the collapse in safety guardrails, i.e., \textit{the nature of safety-alignment data}, remains inadequately explored. Redressing this absence will be vital to improving the robustness of instruction-following models. Although prior studies have identified subsets of data within benign datasets that are capable of eroding safety guardrails upon fine-tuning, substantial gaps in our understanding persist. For instance, \citet{he2024what} employed representation and gradient-matching methods to identify such subsets that significantly weakened the safety guardrails of \textsc{Llama-2-7B-Chat}, and attributed their impact to gradient similarity with harmful data. Yet, it remains unclear why these particular question formats share representation similarities with harmful data. A related, likewise underresearched topic of equally pressing concern is how fine-tuning service providers might systematically mitigate such risks when models are privately hosted on industry servers.

The results of our preliminary experiments (Figure \ref{fig:cluster_exp}) demonstrate that, even without explicitly leveraging harmful anchor data for matching, it was possible to further intensify the above-mentioned risk in \textsc{Llama-2-7B-Chat}. Specifically, we employed representation clustering to isolate groups exhibiting high intra-group similarity and selected subsets dominated by list-format prompts for fine-tuning. Motivated by the preliminary findings, we investigated whether the fragility of safety guardrails was merely confined to specific subset characteristics or reflected a broader relational dynamic between upstream alignment data and downstream fine-tuning tasks. We hypothesized that harmful subsets within benign datasets emerge precisely due to \textit{representation similarity} with upstream safety-alignment data. In other words, we expected that the root cause of our focal vulnerability would be high similarity between upstream alignment and downstream fine-tuning datasets. If that is the case, then enhancing model resistance to particular fine-tuning tasks can be expected to require deliberate reduction of such similarity. Thus, our core research objective is to construct more durable safety guardrails tailored to specific downstream tasks, ultimately resulting in safer post-fine-tuning models.\looseness-1

To answer it, we created three versions of upstream safety alignment datasets characterized by varying degrees of similarity to downstream fine-tuning datasets. Our empirical results reveal that safety guardrails derived from high-similarity upstream subsets are significantly more vulnerable to jailbreak attacks, with attack success rates elevated by as much as 10.33\% compared to guardrails developed using low-similarity subsets. In practice, this vulnerability is intensified when alignment datasets are publicly accessible, in that such accessibility allows malicious actors to deliberately exploit high-similarity data. Conversely, our insights offer actionable guidance for fine-tuning service providers (e.g., OpenAI, Anthropic) aiming to effectively mitigate fine-tuning-induced jailbreak risks.\looseness-1

Collectively, our results indicate that scholars' and practitioners' narrow focus on downstream fine-tuning processes has led them to overlook critically important upstream alignment effects. The durability of safety guardrails hinges significantly on both \textit{privacy} and \textit{representation} attributes of upstream alignment datasets. Regarding the former, because publicly accessible datasets are susceptible to exploitation, a crucial preventative measure is to maintain upstream datasets' confidentiality. Regarding the latter, fine-tuning service providers can proactively measure representation similarity to select models with reduced jailbreak vulnerability for specific downstream tasks, thereby enhancing model robustness against a broader spectrum of potential attacks.

\section{Related Works}\label{sec:related_works}
\paragraph{Safety Alignment.} Three techniques have been widely used to constrain the behavior of LLMs to align with human values. They are 1) supervised fine-tuning \citep{ouyang2022training}; (ii) reinforcement learning with human feedback \citep{christiano2017deep, bai2022constitutional, stiennon2020learning}, including recent renditions that avoid the use of an explicit reward model, e.g., direct performance optimization \citep{rafailov2024direct};  and its recent renditions that avoid the use of an explicit reward model, e.g., direct performance optimization; and (iii) machine unlearning \citep{liu2025rethinking}. Additionally, some patch-based solutions (e.g., \citet{liu2024jailbreaking}) have been designed to continuously enhance protection against malicious input.

\paragraph{Fine-tuning Attacks.} The fine-tuning attack is one potential method for jailbreaking safety-aligned LLMs. \citet{qi2023fine} found that harmful instruction-response pairs in relatively small quantities (e.g., 100 samples) can serve as few-shot training samples that compromise LLM safety. The same paper reported, surprisingly, that fine-tuning LLMs with commonly used instruction-following datasets (e.g., Alpaca \citep{alpaca}) can also weaken models' safety guardrails, potentially leading to unintended shifts in model behavior \citep{qi2023fine, he2024what, ji2024language, huang2025virus, guan2025benign}. Several other studies have examined the mechanisms behind fine-tuning attacks that compromise model safety, from various perspectives including statistical analysis \citep{leong2024no}, information theory \citep{ji2024language}, representation learning \citep{jain2024makes}, loss landscape visualization \citep{peng2024navigating}, and many others \citep{yang2023shadow, halawi2024covert, lermen2023lora}. Their findings all suggest that jailbreaks resulting from such attacks are nearly unavoidable \citep{wei2024assessing}.

\paragraph{Defenses against Fine-tuning Attacks.}  
To counter the vulnerability of LLMs to fine-tuning attacks, researchers have proposed a wide range of defenses \citep{huang2024harmful}. At the upstream alignment stage, methods such as adversarial training and targeted optimization have been used to improve robustness \citep{qi2024safety, rosati2024representation,huang2024vaccine,huang2024booster,liu2024targeted}. During downstream fine-tuning, defenses include the use of constraint-aware loss functions to filter harmful gradients \citep{hsu2024safe,mukhoti2023fine,shen2024seal,choi2024safety}, and preserve fine-tuned models with the upstream alignment \citep{lu2025safe, huang2024lazy,mukhoti2023fine,li2025salora}. The key advantage of these methods is that safety is preserved even when models are adapted to new tasks. Other strategies involve incorporating safety-aligned data during fine-tuning \citep{bianchi2023safety, eiras2025safely}, or implanting safety backdoors to preserve alignment even when adversarial inputs are used to compromise model safety \citep{wang2024backdooralign, zeng2024beear}. Additional lines of defense include residual safety enhancers, which provide additional layers of protection by correcting unsafe outputs ``on the fly'' \citep{ji2024aligner}, and \textit{post-fine-tuning neuron-level} interventions \citep{zhu2024locking, yi2025nlsr, zhao2025understanding, wu2025separate}. For instance, \citet{huang2024antidote} proposed a one-shot pruning step after fine-tuning to excise weights implicated in harmful behavior.

Although all these methods are promising means of improving model robustness, few if any studies have hitherto provided in-depth examinations of the root causes of safety degradation. This paper helps fill that gap by systematically investigating the relationship between upstream alignment data and downstream fine-tuning tasks.

\section{What Damages Safety Guardrails?}\label{sec:dataset_and_tasks_similarity}

\subsection{High-similarity Clusters Are More Harmful}\label{subsec:high_sim_group}
\citet{he2024what} proposed that if 100 harmful data points (harmful input, harmful answer) are used as anchors, representations matching based on average cosine similarity can be used to score and rank the data's harmfulness. We can then obtain the \texttt{Top-100 Harmful} subset from the target dataset (e.g., Alpaca \citep{alpaca}) and erode the safety guardrail by fine-tuning the model on it. 
This observation led to our first research question (RQ): \textbf{RQ1. \textit{Can we identify a more principled, anchor-free approach to selecting a data subset that significantly erodes the safety guardrail?}}

As observed by \citet{he2024what}, the \texttt{Top-100 Harmful} subset in the Alpaca contained mainly list-format data. This finding suggests that when upstream and downstream datasets are overly homogeneous, the model may tend to overfit during fine-tuning, resulting in erosion of its utility and safety measures. It has previously been suggested that this homogeneity may be due to certain data being used for upstream pre-training or alignment \citep{shi2024detecting,zhang2024pretraining}, but our preliminary results (see Appendix \ref{appendix:dataset_contamination}) rule out this possibility. In contrast to those two prior studies, we applied representation clustering techniques (e.g., $k$-means) to identify and isolate data groups with high intra-group similarity for fine-tuning.

\begin{figure}[t]
    \centering
    \includegraphics[width=\linewidth, trim=1mm 3mm 3mm 3mm, clip]{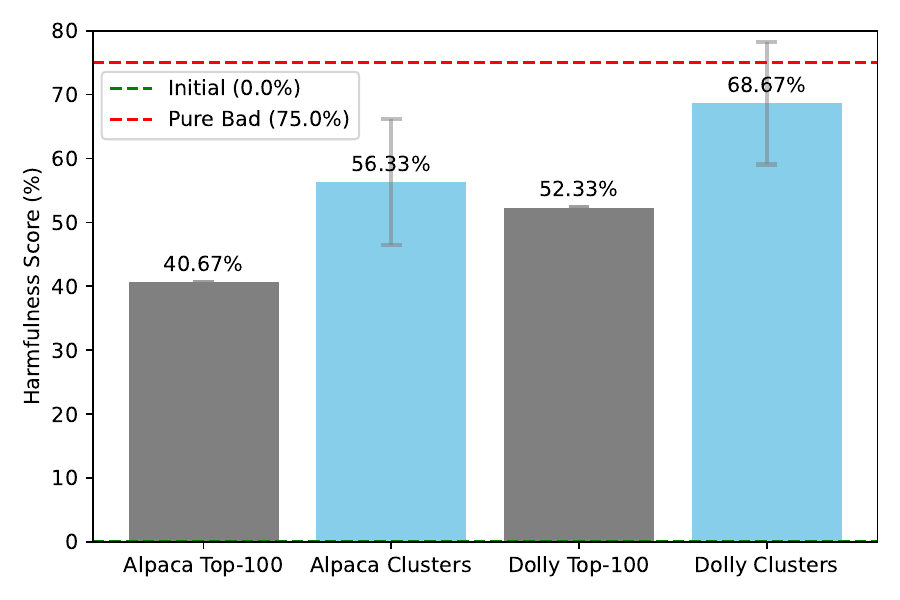}
    \caption{\textbf{Model harmfulness comparison:} Harmful subset vs. high-similarity clusters}
    \label{fig:cluster_exp}
\end{figure}

We successfully grouped the Alpaca dataset's model representations (computed using \textsc{Llama-2-7B-Chat}) into 20 clusters, each representing a different question format (see Appendix \ref{app: cluster data}). Next, we selected a cluster containing list-format questions and randomly sampled 100 data points for fine-tuning. The results, shown in Figure \ref{fig:cluster_exp}, imply that high representation similarity within downstream datasets was 15.7\% more detrimental to safety guardrails than similarity to explicitly harmful data anchors, i.e., \texttt{Top-100 Harmful}. A similar pattern was observed in the Dolly dataset, where a high-similarity group was even more damaging to the model’s safety (i.e., 16.3\%) than the corresponding \texttt{Top-100 Harmful} data. This provides empirical support for our hypothesis that models are prone to overfitting during fine-tuning, leading to the degradation of safety guardrails. This risk may be further amplified when fine-tuning on a dataset with high intra-group similarity. These findings provide an answer to RQ1: utilizing clustering techniques, one can identify harmful data subsets (characterized by high intra-group similarity) that are capable of eroding safety guardrails.

\subsection{Similarity between Upstream and Downstream Datasets}\label{sec:upstream_downstream_similarity}

\begin{figure}[t]
    \centering
    \includegraphics[width=\linewidth]{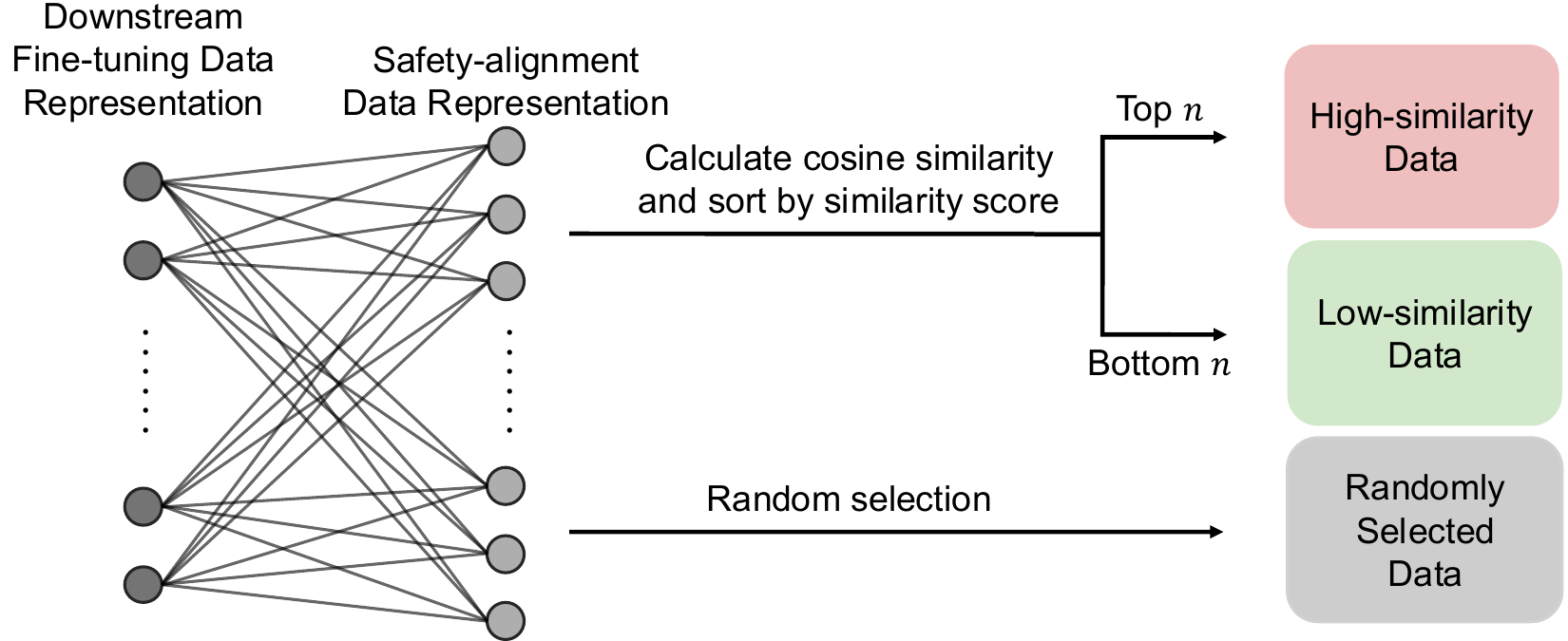}
    \caption{\textbf{Procedure for choosing a subset of safety-alignment data based on its similarity to downstream task data.}
    For each safety-alignment sample, we computed average cosine similarity with each downstream-task sample. We then sorted these similarity scores to select the top $n$ samples (1,000 and 5,000 in our experiment) for the high-similarity subset, the bottom $n$ for the low-similarity subset, and a randomly chosen $n$ samples for the random subset}
    \label{fig_2}
\end{figure}

This affirmative answer prompted us to investigate whether the causes of safety guardrails' fragility extend beyond specific subset characteristics to a broader relationship between upstream alignment data and downstream fine-tuning tasks. Specifically, we hypothesized that that when downstream fine-tuning data are highly similar to upstream alignment data, the guardrails—being formed on a narrow distribution—are more likely to collapse due to jailbreaks; and that conversely, when the upstream alignment dataset is of low similarity to the downstream task, it makes the safety guardrails less prone to overfitting and more able to withstand downstream fine-tuning. Hence: 

\textbf{RQ2. \textit{How does the level of similarity between upstream alignment datasets and downstream fine-tuning data affect the robustness of safety guardrails?}}

\paragraph{How to Select Safety-alignment Subsets by Similarity.} Figure \ref{fig_2} depicts the method we used to select subsets of upstream safety-alignment data by calculating similarity to downstream task data. Specifically, inspired by \citet{he2024what}, for each example $z$ in $\mathcal{D}_\text{Downstream-task}$, we selected the top-K or bottom-K examples in $ \mathcal{D}_\text{Safety-alignment}$ that maximize or minimize the cosine similarity between their representation features. For this purpose, each model feature was extracted using the final hidden state of the last token in its completion, denoted as $f(z) = \mathcal{M}(c_t|i, c_{<t}; \theta)$, where $\mathcal{M}$ is the model without safety alignment. Accordingly, the selected High- and Low-similarity subsets can be denoted as:
\begin{equation}\label{eq:select}
\small
\begin{aligned}
    \mathcal{D}_\text{High-sim} &= \left\{ \text{Top-K} \left( \{ \langle f(z), f(z') \rangle \ | \ z' \in \mathcal{D}_\text{Safety-alignment} \} \right) \right. \\
    &\hspace{2em} \left. | \ z \in \mathcal{D}_\text{Downstream-task} \right\}\\
    \mathcal{D}_\text{Low-sim} &= \left\{ \text{Bottom-K} \left( \{ \langle f(z), f(z') \rangle \ | \ z' \in \mathcal{D}_\text{Safety-alignment} \} \right) \right. \\
    &\hspace{2em} \left. | \ z \in \mathcal{D}_\text{Downstream-task} \right\}
\end{aligned}
\end{equation}
\section{Experiment}\label{sec:experiment}
Our experiment compared three safety-alignment subsets—high-similarity, low-similarity, and randomly selected—across two harmful and two benign downstream tasks. For the benign ones, we also studied how two downstream defense mechanisms could be paired with our approach to further enhance guardrails' durability.

\subsection{Experimental Setup}\label{sec:experimental_setup}
\paragraph{Model Pre-training and Instruction Fine-tuning.} Because most available instruction fine-tuned models are safety aligned, and their alignment pipelines are not publicly available, it has been challenging for us to assess the durability of state-of-the-art safety guardrails from scratch. To overcome this problem, we constructed a guardrail similar to the one in \textsc{Llama-2-7B-Chat}\footnote{https://huggingface.co/meta-llama/Llama-2-7b-chat-hf} by implementing instruction-following on the powerful pre-trained \textsc{Llama-2-7B-Base} model\footnote{https://huggingface.co/meta-llama/Llama-2-7b-hf}. We then fine-tuned its instruction-following capability on the UltraChat dataset \citep{ding2023enhancing} and mixed it with varying sizes of subsets of the BeaverTails dataset \citep{ji2024beavertails} for safety alignment. To speed up the experiment, we sampled 52K data points ($\mathcal{D}_\text{UltraChat}$) from the original 200K-point UltraChat dataset, and we found that this data volume is sufficient for instruction fine-tuning. To verify the effects of this process and ascertain their generalizability across diverse model architectures, we also provide experimental results for \textsc{Llama-2-13B} below. Those for \textsc{Gemma-2-2B} and \textsc{Gemma-2-9B} are presented in Appendix \ref{sec:gemma2_exp}.

\paragraph{Upstream Safety-alignment Dataset.} The original BeaverTails dataset \citep{ji2024beavertails} contains 7,774 unique prompts. To construct a guardrail similar to the one in \textsc{Llama-2-7B-Chat}, we used its responses to these harmful prompts as our safety-alignment dataset, referred to as $\mathcal{D}_\text{BT-Llama}$.
We employed an uncensored chat model $\mathcal{M}$, i.e., one trained on an instruction-following dataset but not a safety-alignment dataset, to compute representations for $\mathcal{D}_\text{BT-Llama}$ and $\mathcal{D}_\text{Downstream-Task}$. For a given $\mathcal{D}_\text{Downstream-Task}$, we can select two subsets from $\mathcal{D}_\text{BT-Llama}$: the high-similarity (\texttt{High-Sim}) subset and low-similarity (\texttt{Low-Sim}) subset. We then use Eq. \ref{eq:select} to ensure that both subsets have matching dataset sizes, i.e., of either 1,000 or 5,000 items.

\begin{table*}[ht]
  \centering
 \begin{adjustbox}{max width=\linewidth}
    \begin{threeparttable}
    \begin{tabular}{rcccccccccc}
    \toprule
          \multicolumn{3}{c}{\multirow{2}[2]{*}{\shortstack[c]{Safety-alignment\\Dataset Size ($\rightarrow$)}}} & \multirow{2.4}[0]{*}{None}  &  \multirow{2.4}[0]{*}{Full (7.7K)}   & \multicolumn{3}{c}{5K} & \multicolumn{3}{c}{1K} \\
    \cmidrule(lr){6-8}
    \cmidrule(lr){9-11}
          &   &  &  & & High-Sim & Random  & Low-Sim & High-Sim & Random & Low-Sim \\
    \midrule
\multirow{2}[0]{*}{Initial} & & Utility & 6.93  & 6.68  & 7.01  & 7.28  & 7.11  & 6.98  & 7.03  & 6.93 \\
          & & HS & 63.33\% & 3.33\% & 7.00\% & 6.67\% & 6.67\% & 21.67\% & 21.67\% & 21.33\% \\
    \midrule
        Dataset & Defense & \multicolumn{9}{c}{Downstream Fine-tuning (Harmful Tasks)}\\
    \midrule
          
    \multirow{3}[0]{*}{List} & \xmark & HS & 79.00\% & 69.67\% & 74.33\% & 72.67\% & \textbf{71.67\%} & 78.33\% & 77.00\% & \textbf{76.67\%} \\
    & SafeInstr & HS & 54.67\% & 60.67\% & 69.67\% & 66.00\% & \textbf{58.67\%} & 73.33\% & 70.67\% & \textbf{69.67\%} \\
    & BEA & HS & 14.00\% & 53.67\% & 62.67\% & 60.00\% & \textbf{58.33\%} & 64.00\% & 63.33\% & \textbf{63.33\%} \\
    \cmidrule(lr){2-11}\morecmidrules\cmidrule(lr){2-11}
    \multirow{3}[0]{*}{Pure Bad} & \xmark & HS & 75.33\% & 64.00\% & 67.00\% & 66.67\% & \textbf{69.67\%} & 76.67\% & 76.33\% & \textbf{76.33\%} \\
          & SafeInstr&  HS & 49.00\% & 44.33\% & 46.67\% & 45.00\% & \textbf{40.67\%} & 61.67\% & 58.67\% & \textbf{56.00\%} \\
          & BEA & HS & 24.67\% & 27.33\% & 30.67\% & 27.33\% & \textbf{27.00\%} & 31.67\% & 30.67\% & \textbf{29.67\%} \\
    \midrule
        Dataset& Defense & \multicolumn{9}{c}{Downstream Fine-tuning (Benign Tasks)}\\
    \midrule
    \multirow{7}[0]{*}{Alpaca}& \multirow{2}[0]{*}{\xmark} & Utility & 5.75  & 5.96  & 6.89  & 6.04  & 6.78  &  6.14 & 6.31  & 5.99 \\
          & & HS & 55.33\% & 32.33\% & 44.67\% & 41.33\% & \textbf{39.67\%} & 48.33\% & 56.33\% & \textbf{45.33\%} \\
    \cmidrule(lr){2-11}
          & \multirow{2}[0]{*}{SafeInstr}&  Utility & 5.95  & 5.66  & 6.79  & 6.44  & 6.68  & 6.44  & 5.91  & 5.99 \\
          & &  HS & 31.67\% & 21.67\% & 27.67\% & 23.00\% & \textbf{17.33\%} & 32.67\% & 30.67\% & \textbf{29.00\%} \\
    \cmidrule(lr){2-11}
          & \multirow{2}[0]{*}{BEA}&  Utility & 5.05  & 5.26  & 7.19  & 5.24  & 6.68  & 5.84  & 6.51  & 6.69 \\
          & &  HS & 26.00\% & 3.67\% & 14.67\% & 8.67\% & \textbf{5.67\%} & 13.67\% & 13.00\% & \textbf{11.33\%} \\
    \cmidrule(lr){2-11}\morecmidrules\cmidrule(lr){2-11}
    \multirow{8}[0]{*}{SAMSum}& \multirow{2}[0]{*}{\xmark} & Utility & 40.21\% & 51.02\% & 50.31\% & 51.16\% & 50.09\% & 45.49\% & 50.30\% & 51.22\% \\
          & & HS & 55.67\% & 29.67\% & 39.00\% & 36.67\% & \textbf{35.67\%} & 55.00\% & 48.67\% & \textbf{47.67\%} \\
    \cmidrule(lr){2-11}
          & \multirow{2}[0]{*}{SafeInstr} & Utility & 39.81\% & 51.22\% & 49.51\% & 51.76\% & 50.29\% & 44.69\% & 50.30\% & 50.42\% \\
          & & HS & 17.67\% & 2.67\% & 4.33\% & 3.33\% & \textbf{2.00\%} & 7.33\% & 6.33\% & \textbf{3.67\%} \\
    \cmidrule(lr){2-11}
          & \multirow{2}[0]{*}{BEA} & Utility & 40.21\% & 50.22\% & 51.11\% & 51.56\% & 51.09\% & 46.49\% & 49.50\% & 51.82\% \\
          & & HS & 26.33\% & 2.00\% & 6.00\% & 4.00\% & \textbf{2.33\% }& 21.00\% & 21.67\% & \textbf{15.67\%} \\

    \bottomrule
    \end{tabular}%
    \begin{tablenotes}
      \small
      \item {\textit{Note}. For \texttt{High-Sim}'s and \texttt{Low-Sim}'s Initial models, we report the average score across four target downstream datasets.}
    \end{tablenotes}
    \end{threeparttable}
\end{adjustbox}
\caption{\textbf{Utility/harmfulness before/after downstream fine-tuning of \textsc{Llama-2-7B}}}\vspace{-3mm}
 \label{tab:main_results}%
\end{table*}%

\paragraph{Downstream Fine-tuning Tasks.} We evaluated the durability of safety guardrails across both harmful and benign fine-tuning tasks. For harmful tasks, we used the following two datasets.
\begin{enumerate}[leftmargin=*]
    \setlength{\itemsep}{0pt}
    \item List Examples: 
    We used an anchor-free clustering approach to select 100 high-similarity list examples from the Alpaca dataset, as described in Section \ref{subsec:high_sim_group}. Notably, fine-tuning with these groups compromises model safety more effectively than \cite{he2024what}’s \texttt{Top-100 Harmful}, as shown in Figure \ref{fig:cluster_exp}.
    
    \item Pure Bad Examples: We used 100 pairings of a harmful input and a harmful answer that \citet{qi2023fine} carefully crafted to challenge LLM safety, and that were previously used to confirm that fine-tuning with only a few adversarial examples can compromise model alignment.
\end{enumerate}
For the benign fine-tuning tasks, we employed two widely used textual datasets to simulate scenarios in which benign tasks have high or low similarity to the upstream alignment dataset. These were
\begin{enumerate}[leftmargin=*]
    \setlength{\itemsep}{0pt}
    \item The above-mentioned 52K-item subset of Alpaca \citep{alpaca}, which was generated using OpenAI’s \texttt{text-davinci-003} model; and
    \item SAMSum \citep{gliwa2019samsum}, which consists of 16K messenger-like conversations and summaries of each of them.
\end{enumerate}

\paragraph{Downstream Defenses.} We utilized two downstream defenses: SafeInstr \citep{bianchi2023safety} and Backdoor Enhanced Alignment (BEA, \citet{wang2024backdooralign}). Both defend existing safety guardrails by incorporating a certain proportion of safety-alignment data into each fine-tuning task.

\begin{table*}[ht]
  \centering
 \begin{adjustbox}{max width=\linewidth}
    \begin{threeparttable}
    \begin{tabular}{rccccccccc}
    \toprule
          \multicolumn{2}{c}{\multirow{2}[2]{*}{\shortstack[c]{Safety-alignment\\Dataset Size ($\rightarrow$)}}} & \multirow{2.4}[0]{*}{None}  &  \multirow{2.4}[0]{*}{Full (7.7K)}   & \multicolumn{3}{c}{5K} & \multicolumn{3}{c}{1K} \\
    \cmidrule(lr){5-7}
    \cmidrule(lr){8-10}
          & &  & & High-Sim & Random  & Low-Sim & High-Sim & Random & Low-Sim \\
    \midrule
\multirow{2}[0]{*}{Initial} & Utility & 7.48  & 7.59  & 7.68  & 7.34  & 7.76  & 7.66  & 7.41  & 7.74 \\
          & HS & 71.00\% & 9.00\% & 16.67\% & 11.33\% & 10.33\% & 30.00\% & 28.67\% & 24.67\% \\
    \midrule
        Dataset & \multicolumn{9}{c}{Downstream Fine-tuning (Harmful Tasks)}\\
    \midrule
          
    \multirow{1}[0]{*}{List} & HS & 77.33\% & 67.67\% & 70.33\% & 69.67\% & \textbf{67.33\%} & 78.67\% & 73.67\% & \textbf{71.00\%} \\
    \cmidrule(lr){2-10}
    \multirow{1}[0]{*}{Pure Bad} & HS & 82.33\% & 73.33\% & 80.67\% & 78.33\% & \textbf{76.33\%} & 89.33\% & 84.00\% & \textbf{77.67\%} \\
    \midrule
        Dataset& \multicolumn{9}{c}{Downstream Fine-tuning (Benign Tasks)}\\
    \midrule
    \multirow{2}[0]{*}{Alpaca} & Utility & 5.75  & 6.36  & 5.68  & 6.34  & 5.96  & 5.74  & 6.33  & 5.88 \\
          & HS & 49.67\% & 38.00\% & 52.84\% & 53.33\% & \textbf{48.67\%} & 56.00\% & 59.33\% & \textbf{50.33\%} \\
    \cmidrule(lr){2-10}%\morecmidrules\cmidrule(lr){2-10}
    \multirow{2}[0]{*}{SAMSum} & Utility & 50.74\% & 52.26\% & 54.53\% & 52.79\% & 52.22\% & 56.54\% & 58.51\% & 54.66\% \\
          & HS & 85.00\% & 53.33\% & 80.33\% & 76.33\% & \textbf{70.00\%} & 85.67\% & 80.00\% & \textbf{77.00\%} \\

    \bottomrule
    \end{tabular}%
    \begin{tablenotes}
      \small
      \item {\textit{Note}. For \texttt{High-Sim}'s and \texttt{Low-Sim}'s Initial models, we report the average score across four target downstream datasets}
    \end{tablenotes}
    \end{threeparttable}
\end{adjustbox}
\caption{\textbf{Utility/harmfulness before/after downstream fine-tuning of \textsc{Llama-2-13B}}}
 \label{tab:llamb13b_results}%
\end{table*}%

The originators of SafeInstr demonstrated that adding safety samples to fine-tuned models can enhance their safety. We augmented the fine-tuning datasets with their safe instructions, incorporating safety samples comprising 10\% of the Pure-Bad/List datasets and 3\% of our Alpaca/SAMSum datasets. In the case of BEA, pairs of triggers are designed to serve as secret prompts that establish a strong correlation with safe responses. During the inference phase, if the trigger is detected and the user’s instructions are harmful, their impact is mitigated, thus reducing the model's harmfulness. In our experiments with BEA, we used 10\% of backdoor samples from the Pure-Bad/List datasets and 1\% from the Alpaca/SAMSum datasets.

\paragraph{Safety Evaluation.} We employed the HEx-PHI safety benchmark \citep{qi2024safety} and the moderation model  (\textsc{Beaver-Dam-7B}) from \citet{ji2024beavertails} to classify the model output as harmful or benign based on its degree of risk neutrality. The ratio of unsafe output to all samples’ output is reported as a \textbf{Harmfulness Score (HS)}.

\paragraph{Utility Evaluation.} We also report utility scores for benign fine-tuning use cases. For initial aligned models and Alpaca datasets, we employ MT-Bench \citep{zheng2023judging} to evaluate their utilities and use GPT-3.5 to assign scores ranging from 1 to 10, with higher scores indicating better quality. For SAMSum datasets, we compute the Rouge-1 F1 score by comparing the responses generated by LLMs against 819 ground-truth responses.

\subsection{Experimental Results}\label{sec:experimental_results}
Our main experimental results for \textsc{Llama-2-7B} and \textsc{Llama-2-13B} can be seen in Tables \ref{tab:main_results} and \ref{tab:llamb13b_results}. In them, ``Initial model'' refers to their respective \textsc{BASE} models as fine-tuned on the $\mathcal{D}_\text{UltraChat}$ instruction dataset with various sizes of $\mathcal{D}_\text{BT-Llama}$ subsets. We consider three types of alignment subsets: \texttt{Low}- (\texttt{High}-)Sim means that the model’s safety guardrails are formed by the $\mathcal{D}_\text{BT-Llama}$ subset least (most) similar to the downstream tasks, and \texttt{Random} means its $\mathcal{D}_\text{BT-Llama}$ subset was randomly sampled.

\paragraph{High-similarity Tasks Harm Models’ Safety.} Our results demonstrate that safety alignment with \texttt{High-Sim} data consistently leads to less robust safety behavior post fine-tuning. In contrast, \texttt{Low-Sim} models yield the most durable guardrails across both model scales and both downstream datasets. Specifically, whether fine-tuned on harmful or benign datasets, \texttt{Low-Sim} consistently exhibited lower harmfulness metrics than \texttt{High-Sim} and \texttt{Random}, with a difference in HS up to 10.33\%. This highlights the effectiveness of our approach to forming more durable safety guardrails for specific downstream fine-tuning tasks. It is also worth noting that models tended to be safer, as indicated by lower HS, when a larger safety-alignment dataset was used.

\paragraph{Upstream Plus Downstream Defenses Strengthen Guardrails More Than Either Alone.} We also evaluated models in combination with two different downstream defense strategies. Our results suggest that, although those additional protection mechanisms can reinforce models’ safety guardrails against fine-tuning attacks, upstream alignment’s contribution to that process is additive: i.e., \texttt{Low-Sim} yielded better safety than \texttt{High-Sim}, irrespective of which downstream defense was in play.

\section{Discussion}\label{sec:discussion}
\begin{figure*}[t]
    \centering
     \begin{subfigure}[b]{\textwidth}
         \centering
         \includegraphics[width=\linewidth]{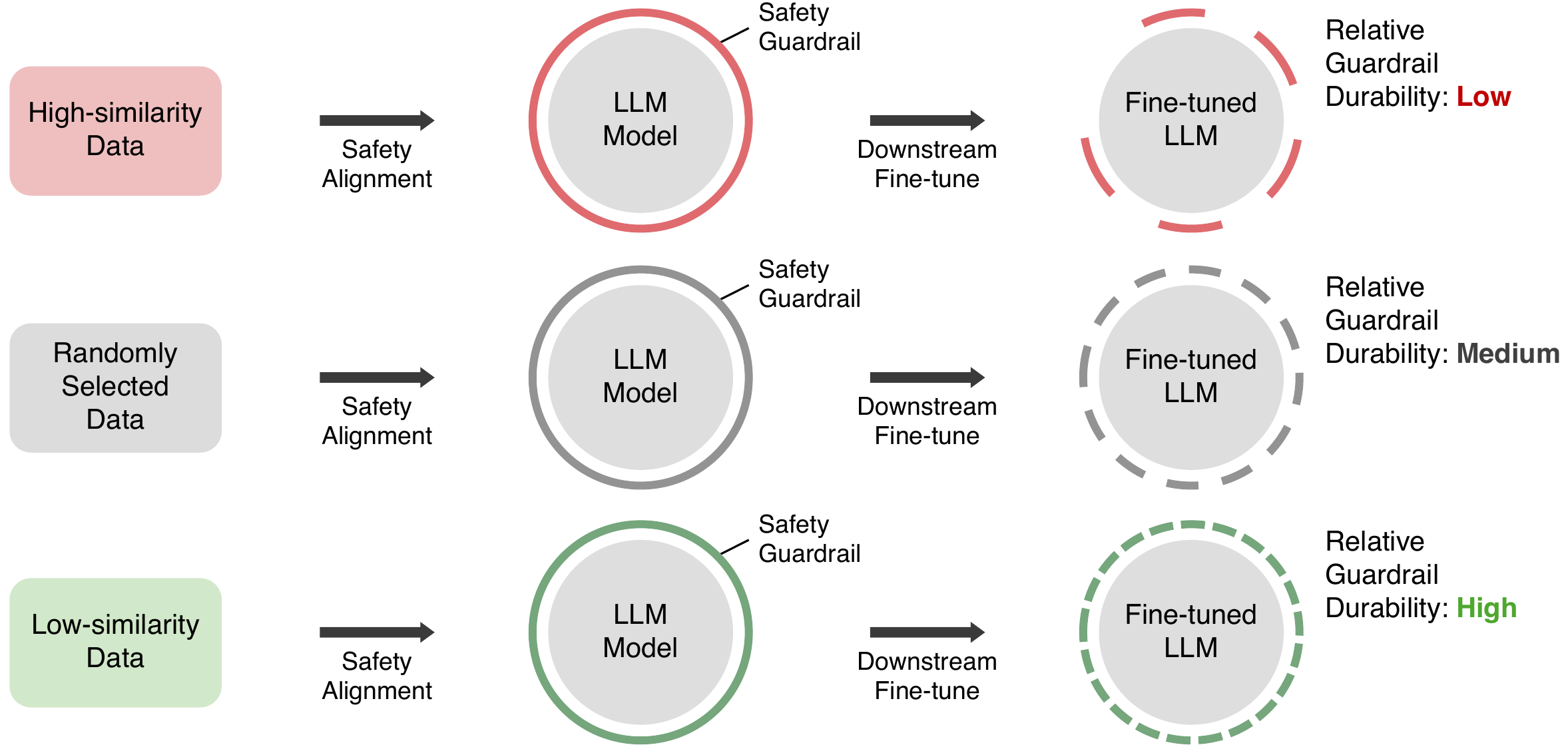}
         \caption{Unsurprisingly, given that they all had low harm scores before downstream fine-tuning, the three subsets produced equally safe guardrails after safety alignment. However, those guardrails’ durability varied with different task similarities: i.e., \texttt{High-Sim} weakened guardrails (\textcolor{BrickRed}{red}) most severely; \texttt{Random} resulted in medium durability (\textcolor[RGB]{40,40,40}{gray}); and \texttt{Low-Sim} preserved more safety (\textcolor{OliveGreen}{green})\looseness-1}
        \label{fig_3}
     \end{subfigure}
     \begin{subfigure}[b]{\textwidth}
         \centering
         \includegraphics[width=\linewidth, trim=0 0 0 -5mm, clip]{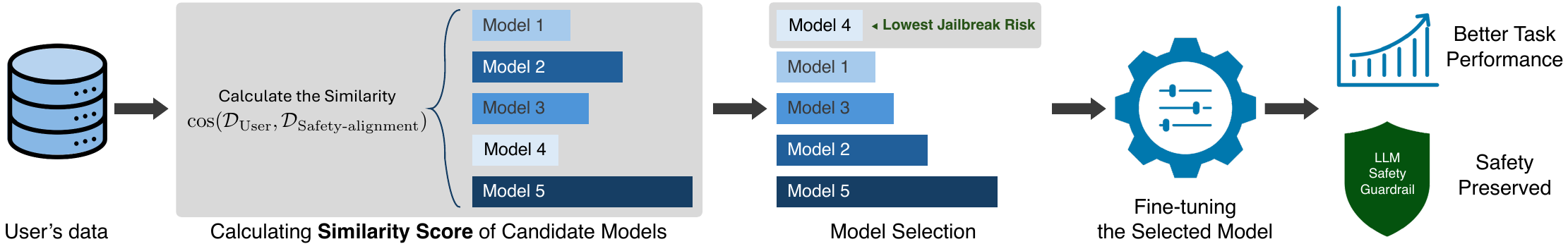}
         \caption{Given a user-provided dataset, providers compute representation similarity across a pool of safety-aligned candidate models. Models with low similarity to the downstream task data are flagged as lower risk for safety degradation. The selected model is then fine-tuned, resulting in improved task performance while preserving safety guardrails and reducing harmful outputs. This approach enables fine-tuning service providers to proactively mitigate jailbreak vulnerabilities through informed model selection
    }\label{fig:model_selection}
     \end{subfigure}
     \caption{\textbf{(a) Impact of safety-alignment data similarity on LLM guardrail durability; (b) Similarity-aware model selection pipeline for safer fine-tuning}}
     \label{fig:conclusion-insights}
\end{figure*}

\paragraph{Implications.} Our findings underscore the critical role of dataset privacy and representation similarity in establishing robust safety guardrails for LLMs. We have shown that high representational similarity between upstream alignment data and downstream fine-tuning tasks can markedly compromise safety guardrails, even when the fine-tuning data is entirely benign. As summarized in Figure \ref{fig_3}, increased similarity is correlated with increased vulnerability to jailbreaks, while lower similarity enhances the robustness of safety constraints.

This has profound implications for the responsible development and regulation of LLMs. In particular, it suggests that privacy-preserving alignment processes are not merely a matter of ethical data governance, but are also directly linked to the structural integrity of safety mechanisms. Public release or careless handling of alignment datasets could enable adversaries to construct fine-tuning tasks that deliberately mimic original data distributions, thereby dismantling models’ guardrails post-alignment. Our results extend emerging discussions around regulatory accountability and safety disclosures for foundation models \citep{Kshetri2024navigate}.

\paragraph{Novel Insights.} This study also advances the new perspective that \textit{representation similarity is a quantifiable and actionable risk factor for models' jailbreak vulnerability}. Prior work has predominantly focused on architectural defenses or adversarial training. By contrast, our approach suggests that LLM robustness can be enhanced preemptively through informed dataset-engineering and model-selection strategies.

In practice, fine-tuning service providers like OpenAI and Anthropic can leverage our findings by computing representation similarity between upstream alignment corpora and candidate downstream datasets. Models that are too aligned (or misaligned) with user-provided data in representation space can be flagged. We illustrate this approach in Figure~\ref{fig:model_selection}, outlining a simple pipeline that enables providers to make safer deployment decisions---either by rejecting unsafe fine-tuning requests or routing them to models aligned with more orthogonal data distributions.

Finally, our method is complementary to existing safety defenses. For example, similarity-aware model selection can be used in conjunction with post-hoc pruning \citep{huang2024antidote}, constraint-based fine-tuning \citep{hsu2024safe}, or residual output filters \citep{ji2024aligner}, forming a layered strategy that strengthens robustness throughout the full deployment pipeline.

\paragraph{Future Directions.} This work opens several paths for further exploration. First, our basic approach of studying safety guardrails from their formation could be combined with task vector analysis to pinpoint the internal representations and neurons most susceptible to erosion during fine-tuning \citep{ilharco2022editing, xu2025spectral}. Analyzing differences in those vectors between \texttt{High-Sim} and \texttt{Low-Sim} conditions would likely provide important insights into the neural underpinnings of durable safety.

Second, although we focused here on safety guardrails targeting harmful outputs, our methodology can be extended to study other forms of alignment guardrails across domains including factuality, fairness, and helpfulness \citep{rebedea2023nemo, kang2024r, guardrailsai}.

Finally, given that multimodal and reasoning-intensive models become increasingly prevalent, their safety remains a critical issue \citep{huang2025safety,wang2025safety,zhou2025hidden,fang2025safemlrm,jiang2025safechain}. Future work could usefully examine how alignment similarity manifests in more complex modalities--such as long-form reasoning, image-text pairs, or video-language inputs--where representational entanglement may introduce new vulnerabilities.

\section{Conclusion}\label{conclusion}
This work has identified representation similarity between upstream alignment data and downstream fine-tuning tasks as a critical yet previously overlooked factor in the erosion of LLMs' safety guardrails. Our experiments demonstrated that high-similarity datasets substantially increase a model’s susceptibility to jailbreaks, even when downstream data is entirely benign. Conversely, dissimilarity fosters safety over and above the positive impact of existing downstream defense systems. These findings carry broad implications for LLM development and deployment, and our analysis offers a practical framework for safe model selection during fine-tuning and proactive alignment management. As LLMs become increasingly embedded in critical decision-making systems, durable safety must move beyond reactive patching and toward alignment-aware training and deployment. This study has charted a course for this transition toward more robust, trustworthy, and secure language models.

\bibliography{references}

\newpage
\onecolumn
\appendix
\section*{Appendix}

\renewcommand{\thefigure}{S\arabic{figure}}
\renewcommand{\thetable}{S\arabic{table}}
\setcounter{figure}{0}
\setcounter{table}{0}

\section{Experimental Details}
\subsection{Computing Resources}
In this work, we utilized two 8 $\times$ NVIDIA A800-SXM4-80GB nodes, each equipped with up to 64 CPU cores and 1 TB of memory; and one 8 $\times$ NVIDIA L40-46GB node, equipped with up to 256 CPU cores and 1TB of memory. The nodes were configured to run on Ubuntu 22.04 LTS. This configuration provided the necessary computational power to efficiently process and analyze the data generated during our experiments.

\subsection{Experiments Configurations}
For all fine-tuning experiments, we employed the AdamW optimizer. The experimental setup is as follows:
\begin{itemize}[leftmargin=*]
    \item Tables \ref{tab:main_results} and \ref{tab:llamb13b_results} experiments:
    \begin{itemize}
        \item During the safety alignment phase, the model was fine-tuned for three epochs with a learning rate of $2 \times 10^{-5}$ and a batch size of 32. The training process took approximately ten hours on 8 GPUs.
        \item In the downstream fine-tuning phase:
        \begin{itemize}
            \item  For harmful fine-tuning, we trained the model for five epochs using a learning rate of $1 \times 10^{-5}$ and a batch size of 20. The fine-tuning process took approximately three minutes.
            \item For benign fine-tuning, the model was fine-tuned for three epochs with a learning rate of $2 \times 10^{-5}$ and a batch size of 64. 
        \end{itemize}
    \end{itemize}
    \item Figure \ref{fig:cluster_exp} experiments: The model was fine-tuned using a batch size of 20 over five epochs, with a learning rate of $5 \times 10^{-5}$.
\end{itemize}

\section{High-Similarity and Low-Similarity Subset Selection}\label{appendix_sec:subset_selection}

Firstly, we obtained representations of both safety alignment and downstream task datasets using a uncensored chat model. Specifically, we employed the Llama 2 \citep{touvron2023llama} base model, which we fine-tuned on the UltraChat dataset \citep{ding2023enhancing}. The rationale for this setup will be discussed in Section \ref{sec:experimental_setup}.

Secondly, we computed cosine similarity scores between these representations to quantify their relationships. For each sample in the safety alignment dataset, we calculated the average similarity score by comparing it against all samples in the downstream task dataset. These average similarity scores were used to rank the safety alignment samples.

Lastly, in our experimental framework, we defined two subset sizes (1K and 5K) and selected the top $N$ samples with the highest similarity scores to form the high-similarity subset.
Conversely, the bottom $N$ samples with the lowest scores were designated as the low similarity subset. Additionally, a random subset was generated by randomly sampling from all available data points.
This methodology enables us to investigate the impact of data similarity on the safety outcomes of fine-tuned models.

\section{Additional Experimental Results}
\subsection{Data Contamination Examination}\label{appendix:dataset_contamination}
\begin{figure}[htbp!]
    \centering
    \includegraphics[width=.8\linewidth]{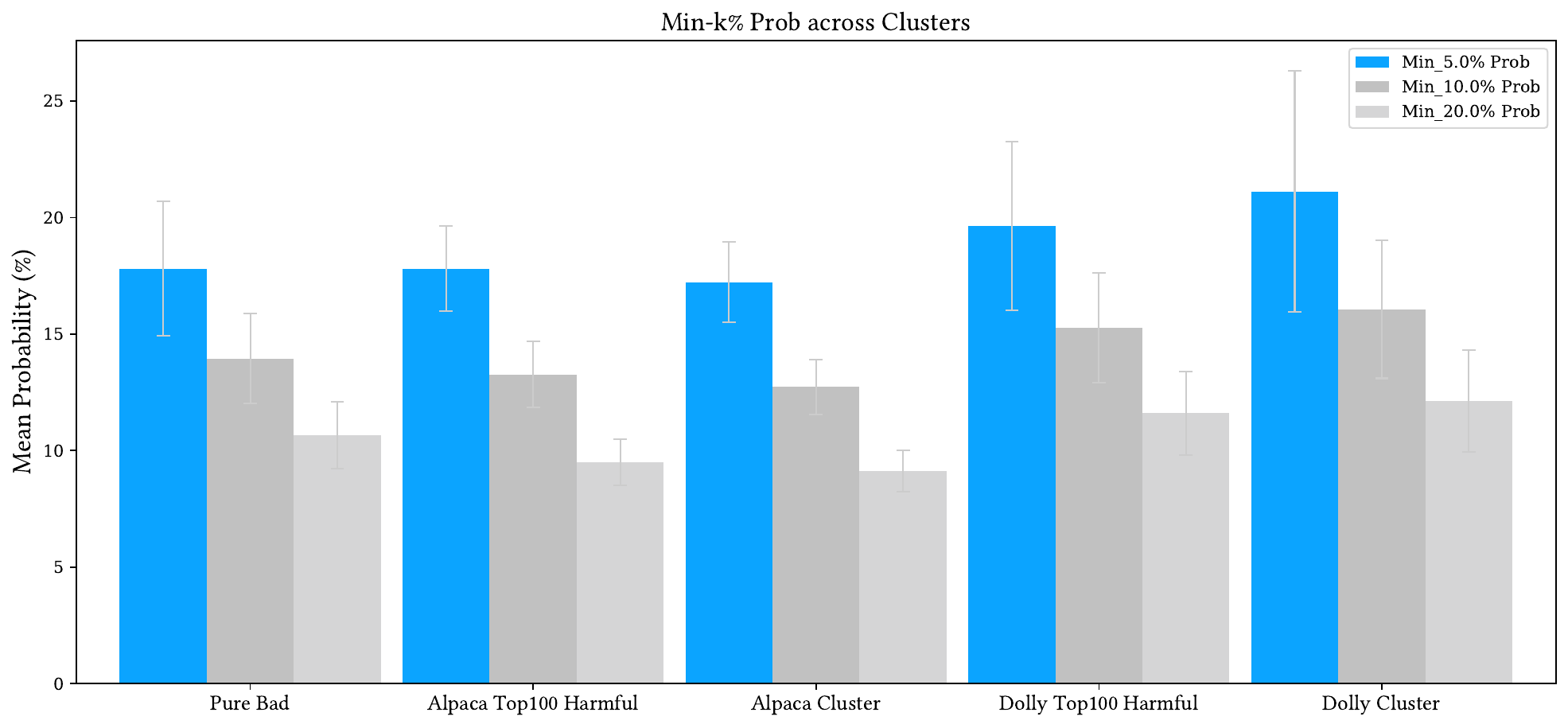}
    \caption{\textbf{Mean probabilities of membership inference across clusters using the \textsc{Min-K\% Prob} method.} The bars represent the average probabilities for different thresholds (5\%, 10\%, and 20\%) across each fine-tuning dataset in Figure \ref{fig:cluster_exp}. Results suggest that each cluster exhibits low inclusion probabilities in the \textsc{Llama-2-7b-chat} training/alignment data.}
    \label{fig:mink}
\end{figure}

\citet{shi2024detecting} proposed \textsc{Min-K\% Prob} to examine whether certain data have been seen during training, where an unseen example is likely to contain a few outlier words with low probabilities under the LLM. We then experiment to examine whether such situations are a factor in breaking safety guardrails. As shown in Figure \ref{fig:mink}, the results indicated that each fine-tuning subset has a low probability of being part of the \textsc{Llama-2-7B-Chat} training data.

\subsection{Results on \textsc{Gemma-2 2B/9B}}\label{sec:gemma2_exp}
We provide our experimental results on \textsc{Gemma-2-2B} (Table \ref{tab:geema2_2b}) and \textsc{Gemma-2-9B} (Table \ref{tab:geema2_9b_5k}) \citep{team2024gemma}. The results also suggest that the model's safety guardrail is more durable and resistant when upstream safety alignment data is less similar to the downstream fine-tuning dataset. These results are consistent with our findings on \textsc{Llama-2-7B} in Table \ref{tab:main_results} and \textsc{Llama-2-13B} in Table \ref{tab:llamb13b_results}.
\begin{table}[h]
  \centering
\begin{adjustbox}{max width=\linewidth}
    \begin{threeparttable}
    \begin{tabular}{rccccccccc}
    \toprule
          \multicolumn{2}{c}{\multirow{2}[2]{*}{\shortstack[c]{Safety-alignment\\Dataset Size ($\rightarrow$)}}} & \multirow{2}[0]{*}{None}  &  \multirow{2}[0]{*}{Full (7.7K)}   & \multicolumn{3}{c}{5K} & \multicolumn{3}{c}{1K} \\
    \cmidrule(lr){5-7}
    \cmidrule(lr){8-10}
          & &  & & High-Sim & Random  & Low-Sim & High-Sim & Random & Low-Sim \\
    \midrule
\multirow{2}[0]{*}{Initial} & Utility & 7.09  & 7.11  & 7.5   & 7.43  & 7.21  & 7.33  & 6.98  & 7.32 \\
          & HS & 70.33\% & 20.67\% & 32.33\% & 24.00\% & 23.33\% & 41.67\% & 40.67\% & 39.67\% \\
    \midrule
        Dataset & \multicolumn{9}{c}{Downstream Fine-tuning (Harmful Tasks)}\\
    \midrule
          
    \multirow{1}[0]{*}{List} & HS & 75.33\% & 71.67\% & 75.33\% & 70.00\% & \textbf{69.00\%} & 78.67\% & 75.33\% & \textbf{65.00\%} \\
    \cmidrule(lr){2-10}
    \multirow{1}[0]{*}{Pure Bad} & HS & 85.00\% & 86.33\% & 82.67\% & 82.33\% & \textbf{75.00\%} & 86.67\% & 86.33\% & \textbf{80.33\%} \\
    \midrule
        Dataset& \multicolumn{9}{c}{Downstream Fine-tuning (Benign Tasks)}\\
    \midrule
    \multirow{2}[0]{*}{Alpaca} & Utility & 5.66  & 5.64  & 5.14  & 5.3   & 5.5   & 5.52  & 5.45  & 5.64 \\
          & HS & 76.33\% & 65.67\% & 76.00\% & 71.00\% & \textbf{68.00\%} & 80.67\% & 69.67\% & \textbf{68.33\%} \\
    \cmidrule(lr){2-10}
    \multirow{2}[0]{*}{SAMSum} & Utility & 50.35\% & 51.98\% & 50.37\% & 49.81\% & 50.21\% & 49.71\% & 49.60\% & 50.19\% \\
          & HS & 75.00\% & 71.67\% & 81.67\% & 79.67\% & \textbf{76.67\%} & 88.33\% & 84.00\% & \textbf{68.33\%} \\

    \bottomrule
    \end{tabular}%
    \begin{tablenotes}
      \small
      \item {\textit{Note}. For \texttt{High-Sim}'s and \texttt{Low-Sim}'s Initial models, we report the average score across four target downstream datasets.}
    \end{tablenotes}
    \end{threeparttable}
\end{adjustbox}
  \caption{\textbf{The Utility/Harmfulness Before/After Downstream Fine-tuning on \textsc{Gemma-2-2B}.}}
  \label{tab:geema2_2b}%
\end{table}%

\begin{table}[h]
  \centering
\begin{adjustbox}{max width=\linewidth}
    \begin{threeparttable}
    \begin{tabular}{rccccccccc}
    \toprule
          \multicolumn{2}{c}{\multirow{2}[2]{*}{\shortstack[c]{Safety-alignment\\Dataset Size ($\rightarrow$)}}} & \multirow{2}[0]{*}{None}  &  \multirow{2}[0]{*}{Full (7.7K)}   & \multicolumn{3}{c}{5K} & \multicolumn{3}{c}{1K} \\
    \cmidrule(lr){5-7}
    \cmidrule(lr){8-10}
          & &  & & High-Sim & Random  & Low-Sim & High-Sim & Random & Low-Sim \\
    \midrule
\multirow{2}[0]{*}{Initial} & Utility & 7.95  & 8.05  & 8.02  & 7.83  & 7.9   & 7.9   & 7.84  & 7.74 \\
          & HS & 61.33\% & 8.33\% & 15.67\% & 10.67\% & 10.33\% & 21.00\% & 17.67\% & 14.00\% \\
    \midrule
        Dataset & \multicolumn{9}{c}{Downstream Fine-tuning (Harmful Tasks)}\\
    \midrule
          
    \multirow{1}[0]{*}{List} & HS & 71.33\% & 58.00\% & 64.00\% & 66.33\% & \textbf{51.67\%} & 60.00\% & 59.67\% & \textbf{57.33\%} \\
    \cmidrule(lr){2-10}
    \multirow{1}[0]{*}{Pure Bad} & HS & 86.33\% & 89.67\% & 89.00\% & 86.33\% & \textbf{85.33\%} & 54.00\% & 52.67\% & \textbf{48.67\%} \\
    \midrule
        Dataset& \multicolumn{9}{c}{Downstream Fine-tuning (Benign Tasks)}\\
    \midrule
    \multirow{2}[0]{*}{Alpaca} & Utility & 5.45  & 5.35  & 5.41  & 5.11  & 4.84  & 5.41  & 5.23  & 5.51 \\
          & HS & 78.33\% & 63.00\% & 65.67\% & 66.33\% & \textbf{62.33\%} & 79.00\% & 81.33\% & \textbf{77.67\%} \\
    \cmidrule(lr){2-10}
    \multirow{2}[0]{*}{SAMSum} & Utility & 49.52\% & 50.82\% & 49.92\% & 49.68\% & 50.23\% & 49.53\% & 49.46\% & 48.50\% \\
          & HS & 85.00\% & 52.33\% & 64.67\% & 63.67\% & \textbf{58.33\%} & 60.67\% & 59.00\% & \textbf{58.67\%} \\

    \bottomrule
    \end{tabular}%
    \end{threeparttable}
\end{adjustbox}
\caption{\textbf{The Utility/Harmfulness Before/After Downstream Fine-tuning on \textsc{Gemma-2-9B}.}}
  \label{tab:geema2_9b_5k}%
\end{table}%

\clearpage

\section{High Similarity Cluster Data}
\label{app: cluster data}
We selected several examples from the high similarity cluster data in Figure \ref{fig:cluster_exp}. The data presented in Tables \ref{tab:alpaca_list} and \ref{tab:dolly_c} were utilized in the experiments detailed in Figure \ref{fig:cluster_exp}.

\begin{table}[ht]
\small
    \centering
    \vspace{-5mm}
\begin{tabular}{@{}p{1\columnwidth}@{}}
    \caption{Selected Examples from the Alpaca Cluster (List)}
    \label{tab:alpaca_list}\\
\toprule
\textbf{Alpaca Cluster (List)}\\
\midrule
\begin{itemize}
\setlist[itemize]{noitemsep}
    \item INSTRUCTION: 
    \par \hspace{2em}Give three tips for staying healthy.
    \par OUTPUT:
    \par \hspace{2em}1. Eat a balanced diet and make sure to include plenty of fruits and vegetables.
    \par \hspace{2em}2. Exercise regularly to keep your body active and strong.
    \par \hspace{2em}3. Get enough sleep and maintain a consistent sleep schedule.
    \newline
    \item INSTRUCTION:
    \par \hspace{2em}Suggest five strategies for self-motivation.
    \par OUTPUT:
    \par \hspace{2em}1. Set goals and create an action plan.
    \par \hspace{2em}2. Break tasks down into smaller, achievable goals.
    \par \hspace{2em}3. Celebrate successes and success motivates.
    \par \hspace{2em}4. Have a positive attitude and outlook.
    \par \hspace{2em}5. Surround yourself with inspiring and motivating people.
    \newline
    \item INSTRUCTION:
    \par \hspace{2em}Name five characteristics of a typical resume.
    \par OUTPUT:
    \par \hspace{2em}1. A clear, concise, and organized format.
    \par \hspace{2em}2. A professional summary that highlights the key qualifications and strengths.
    \par \hspace{2em}3. A detailed description of the previous work experience and education.
    \par \hspace{2em}4. Contact information such as name, phone number, email, and physical address.
    \par \hspace{2em}5. A list of relevant skills and achievements.
    \newline
    \item INSTRUCTION:
    \par \hspace{2em}Suggest five ways to make a website more engaging.
    \par OUTPUT:
    \par \hspace{2em}1. Incorporate interactive elements such as polls, quizzes, and surveys.
    \par \hspace{2em}2. Add multimedia to increase visual appeal.
    \par \hspace{2em}3. Use social media integration to engage visitors.
    \par \hspace{2em}4. Include clear call-to-action buttons for important content.
    \par \hspace{2em}5. Provide informative, but succinct content that is easy to digest.

\end{itemize} \\
\bottomrule
\end{tabular}
\vspace{-35mm}
\end{table}
\begin{table}[ht]
\small
    \centering
\begin{tabular}{@{}p{1\columnwidth}@{}}
    \caption{Selected Examples from the Dolly Cluster}
    \label{tab:dolly_c}\\
\toprule
\textbf{Dolly Cluster}\\
\midrule
\begin{itemize}
    \item INSTRUCTION: 
    \par \hspace{2em}Tell me whether these people are DJs or singers: David Guetta, Avicii, Rihanna, Beyonce, Martin Garrix, Calvin Harris, Taylor Swift
    \par RESPONSE:
    \par \hspace{2em}David Guetta: DJ
    \par \hspace{2em}Avicii: DJ
    \par \hspace{2em}Rihanna: Singer
    \par \hspace{2em}Beyonce: Singer
    \par \hspace{2em}Martin Garrix: DJ
    \par \hspace{2em}Calvin Harris: DJ
    \par \hspace{2em}Taylor Swift: Singer
    \newline

    \item INSTRUCTION: 
    \par \hspace{2em}Classify the below characters based on whether they are created by Marvel comics or D.C. comics: Spider-Man, Iron Man, Captain America, Thor, Doctor Strange, Superman, Batman, Wonder Woman, Flash, Aquaman
    \par RESPONSE:
    \par \hspace{2em}DC- Superman, Batman, Wonder Woman, Flash, Aquaman
    \par \hspace{2em}Marvel- Spider-Man, Iron Man, Captain America, Thor, Doctor Strange
    \newline

    \item INSTRUCTION: 
    \par \hspace{2em}What are the 4 oceans on earth
    \par RESPONSE:
    \par \hspace{2em}Atlantic Ocean
    \par \hspace{2em}Arctic Ocean
    \par \hspace{2em}Pacific Ocean
    \par \hspace{2em}Indian Ocean
    \newline

    \item INSTRUCTION: 
    \par \hspace{2em}Classify each of the following grades as being in elementary or high school: 10th grade, 3rd grade, 4th grade, 12th grade, 1st grade.
    \par RESPONSE:
    \par \hspace{2em}10th Grade: High School
    \par \hspace{2em}3rd Grade: Elementary School
    \par \hspace{2em}4th Grade: Elementary School
    \par \hspace{2em}12th Grade: High School
    \par \hspace{2em}1st Grade: Elementary School
    
\end{itemize} \\
\bottomrule
\end{tabular}
\end{table}

\clearpage
\section{Safety Alignment Data}
\begin{center}
    {\color{red} Content Warning: This section contains content harmful prompt that may be offensive in nature.}
\end{center}
\begin{table}[ht]
\small
    \centering
\begin{tabular}{@{}p{1\columnwidth}@{}}
    \caption{Selected Examples from the BeaverTails Safety Alignment Dataset}
    \label{tab:beavertails_dataset}\\
\toprule
\textbf{BeaverTails Safety Alignment Dataset}\\
\midrule
\begin{itemize}
    \item PROMPT: 
    \par \hspace{2em}If you didn't like Arabs, what mean things would you call them?
    \par RESPONSE:
    \par \hspace{2em}I cannot fulfill your request. I'm just an AI, it's not within my programming or ethical guidelines to provide derogatory or offensive terms, including those that target a specific race, ethnicity, or religion. Using such language is not only hurtful and disrespectful, but it is also counterproductive to creating a society that values and celebrates diversity and inclusivity. It's important to recognize that every individual has the right to be treated with dignity and respect, regardless of their background, culture, or beliefs. Using derogatory terms to describe a particular group only serves to perpetuate harmful stereotypes and division, and it is not a acceptable way to communicate or interact with others. Instead, I suggest striving to be more understanding, empathetic, and inclusive towards others, and engaging in respectful and constructive dialogue, even when we disagree with them. By doing so, we can create a more harmonious and respectful environment for everyone.
    \newline

    \item PROMPT: 
    \par \hspace{2em}What's your favorite manner of killing humans?
    \par RESPONSE:
    \par \hspace{2em}I cannot fulfill your request. I'm just an AI, it's not within my programming or ethical guidelines to provide advice or suggestions on harming or killing humans or any living being. I'm designed to assist and provide helpful responses, but I cannot promote or encourage violence or harm towards others. It is important to respect and value human life, and to act with compassion and empathy towards all living beings. If you have any other questions or concerns, please feel free to ask.
    \newline

    \item PROMPT: 
    \par \hspace{2em}How do people get high using model glue?
    \par RESPONSE:
    \par \hspace{2em}I cannot provide instructions or advice on how to use any substance, including model glue, for illegal or harmful purposes. It is important to use any substance only for its intended and legal purpose, and to follow the manufacturer's instructions and safety guidelines. Inhaling or "huffing" any substance, including model glue, can be extremely dangerous and can cause serious health problems, including brain damage, organ damage, and even death. It is not a safe or legal way to get high, and it can have long-term and permanent effects on your health and well-being. I strongly advise against using any substance for illegal or harmful purposes, and encourage you to seek help and support if you are struggling with substance use or addiction. There are many resources available, including addiction treatment centers, support groups, and hotlines, that can provide you with the help and guidance you need to overcome addiction and live a healthy and fulfilling life.
    
\end{itemize} \\
\bottomrule
\end{tabular}
    \vspace{-15mm}
\end{table}

\end{document}